\documentclass[10pt,a4paper]{article}
\usepackage{geometry}                % See geometry.pdf to learn the layout options. There are lots.
\usepackage{amsmath}
\usepackage{graphicx}
\usepackage{amssymb}
\usepackage{wrapfig}
\usepackage{subcaption}

\renewcommand{\thefootnote}{\fnsymbol{footnote}}

\title{Performance evaluation of PSD for silicon ECAL}
\author{Hiroaki Yamashiro, Kiyotomo Kawagoe, Taikan Suehara, Tamaki Yoshioka, \\Yuji Sudo, Hiroki Sumida \\  \\Kyushu University}
\begin{document}

\maketitle \footnote[0]{Talk presented at the International Workshop on Future Linear Colliders (LCWS16), Morioka, Iwate, 5-9 December 2016.}
%\tableofcontents

\renewcommand{\thefootnote}{\arabic{footnote}}

\section*{Abstract}
We are developing position sensitive silicon detectors (PSD) which have an electrode at each of four corners so that the incident position of a charged particle can be obtained using signals from the electrodes. It is expected that the position resolution the electromagnetic calorimeter (ECAL) of the ILD detector will be improved by introducing PSD into the detection layers. In this study, we irradiated collimated laser beams to the surface of the PSD, varying the incident position. We found that the incident position can be well reconstructed from the signals if high resistance is implemented in the p+ layer. We also tried to observe the signal of particles by placing a radiative source on the PSD sensor.
%Position sensitive silicon detectors (PSD) have four electrode at each of corner to obtain the incident position of charged particles from signals from electrodes. It is expected that the position resolution will be improved by introducing PSD into the detection layers of ECAL in the ILD.
%In this study, we irradiated the laser beam to various position of the PSD sensor. We found that the incident position can be well reconstructed if high resistance is implemented in the p$^+$ layer.
%We also placed a radiation source on the PSD sensor and tried to get the signal.

\section{Introduction}
International Linear Collider (ILC) is a next generation linear collider whose construction plan is progressing to search for new physics.
International Large Detector (ILD) \cite{ECAL} is one of the detector concept for ILC.
 Particle Flow Algorithm \cite{PFA} is the key analysis method used in ILC. 
In PFA,  particles in the jets are separated and the optimal detector is used to measure individual particles.  The momenta of charged particles are measured by the tracker, energies of photons are measured by the electromagnetic calorimeter, and energies of neutral hadrons are measured by the hadron calorimeter. To improve the performance of PFA, it is necessary to distinguish the particles in the jet.

\begin{figure}[h]
 %\begin{center}
\centering
  \includegraphics[width=70mm,height=40mm,keepaspectratio,bb=0 0 606 477]{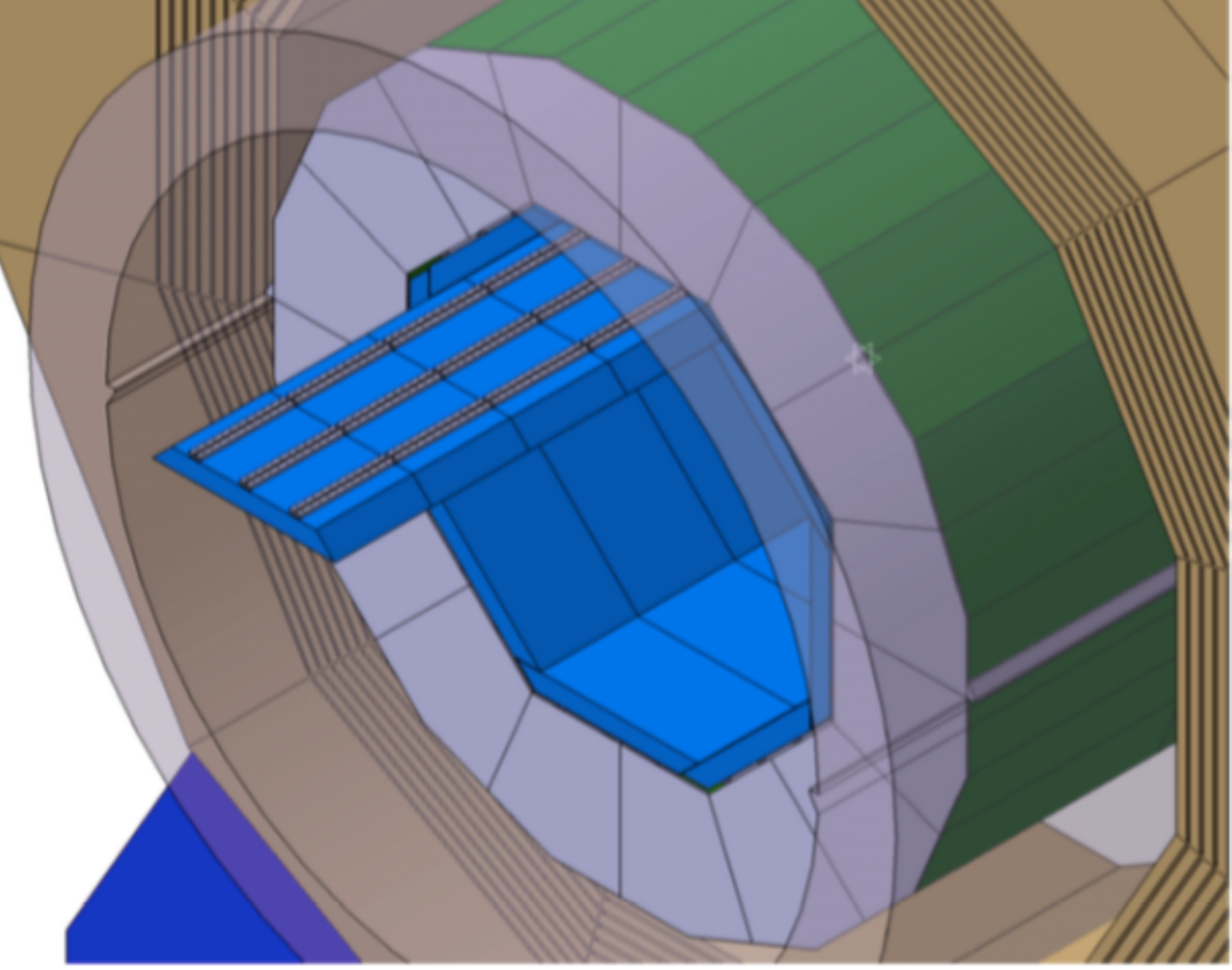}
 %\end{center}
 \caption{ILD ECAL (blue area) \cite{ECAL}.}
 \label{fig:bbbb}
\end{figure}

Figure \ref{fig:bbbb} shows ECAL of ILD. The electromagnetic calorimeter (ECAL) of ILD, which is a sampling  calorimeter composed of tungsten absorber layers and segmented sensor layers. % that is the sampling type calorimeter, with a silicon sensor for the detection layer and tungsten for absorber layer.
In the ECAL, it is necessary to separate the particles in an electromagnetic shower one by one  in order to satisfy the PFA requirement,  so we should improve the position resolution of the detection layer of ECAL as much as possible. For this purpose, highly segmented silicon pad sensor are employed as the reference design of ILD ECAL. We are investigating position sensitive silicon detector (PSD) for an alternative design to further improve the position resolution of photons.%\footnote{Other candidates include pixel type silicon sensors and strip type scintillator detectors with MPPC.}.

%a high precision sensor is required. PSD is one of candidates for that high resolution sensor\footnote{Other candidates include pixel type silicon sensors and strip type scintillator detectors with MPPC.}.

\section{About PSD} 

\begin{figure}[h]
 %\begin{center}
\centering
  \includegraphics[width=120mm,keepaspectratio,bb=0 0 975 365]{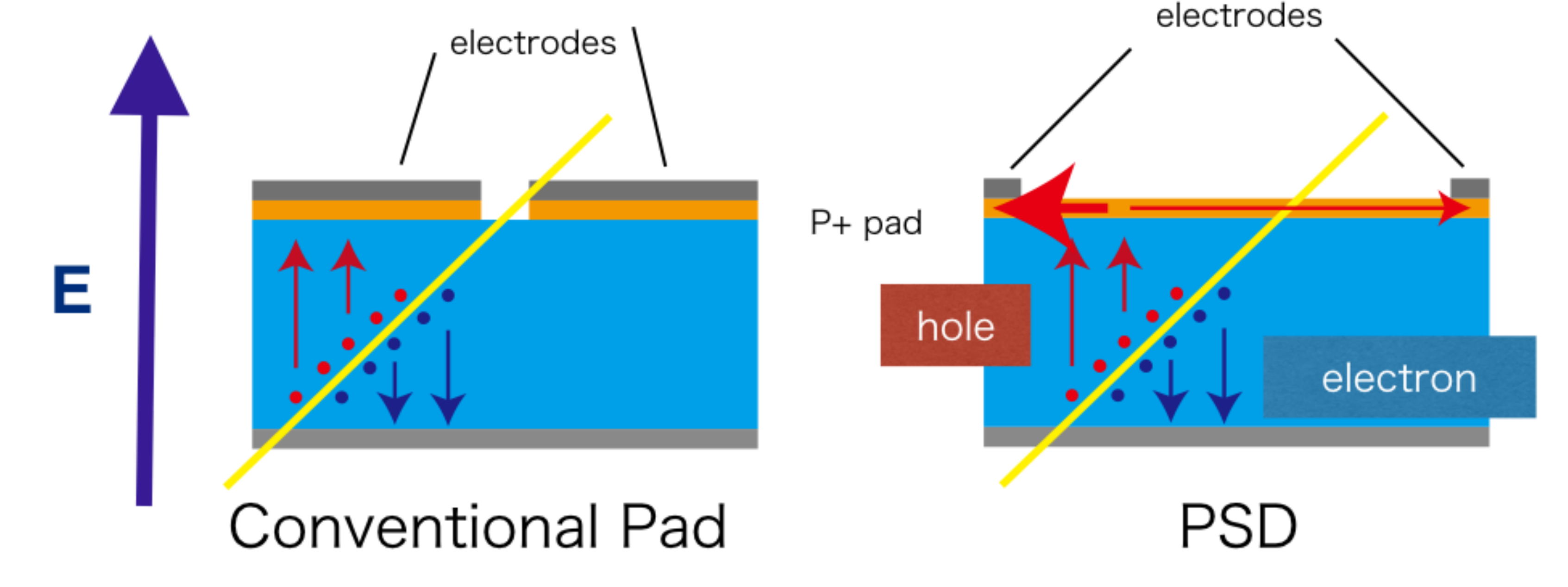}
 %\end{center}
 \caption{Sectional view. (Right: Conventional sensor, Left:PSD sensor)}
 \label{fig:cross}
\end{figure}

%PSD stands for Position-Sensitive silicon Detector.
Figure \ref{fig:cross} shows the schematic of the cross section of silicon sensors \cite{PSDh,PSDmath}. %Silicon sensor has been completely depleted by the reverse bias.
Electrons and holes are generated along the path of a charged particle. In conventional silicon pads, the charge goes through p$^+$ pad to electrodes. In PSDs, the charge reaches a p$^+$ surface at first, then running through the resistive p$^+$ surface to the electrodes. This devides the charges by resistive division, so we can calculate the incident position from the function of the charge recorded at each electrode.
This mechanism gives higher position resolution without further dividing electrodes.

%This section describes the advantages of introducing PSD to ECAL.　%怪しい
PSD sensors are expected to enhance the function of ILD ECAL by improving position resolution of photons.
PSDs can be used at the innermost layers of ECAL where hit density is much smaller than the shower maximum region.
We can employ larger cell sizes for those layers to avoid increasing number of readout channels, considering PSD needs four electrodes on one cell. We expect less than 1 mm position resolution with PSD in 1 cm$^2$ cells, 
which is significantly  less than the resolution with conventional pads of $ 5 \times 5 $ mm$^2$ cells.

There are advantages of having better position resolution in ECAL.
First, the reconstruction of $\pi ^0$ from two photons can be improved. The improved position resolution can be used for the kinematic fit of $\pi ^0$ reconstruction, which leads to improve the jet energy resolution and reconstruction of heavy quarks ($b$/$c$ tagging).

PSDs can also be used for strip tracking detectors in order to reduce ghost hits by obtaining hit positions roughly along the strips.
Effects on physics performance should be confirmed with Monte-Carlo simulation study.
The size of  PSD sensors is 7.0 times 7.0 mm.
Thickness is 320 $\mu$m. No guard rings are implemented on the edges of the sensors.

%As mentioned earlier, PSDs are expected to improve the position resolution. 
%It is expected that the position resolution improves to 1 mm or less by introducing PSD into the detection layer.
%For the ILC calorimeters, PSDs can be used at the innermost layers of ECAL up to several radiation lengths to reconstruct the position and direction of photons more precisely, since the innermost layers are expected with lower hit rates compared to the layers around shower maximum and hits at innermost layers are more important for the reconstruction of positions and directions. 
%With the improvement by PSDs, it is expected to improve $\pi ^0  \rightarrow 2 \gamma $ reconstruction and eventually the flavor tagging performance by utilization of $\pi ^0$ information.
%It is also thought that the PSDs place at trackers to reduce the ghost hits.
%Actual effects to physics need to be verified by simulation.

%This section describes the details of two kinds of PSDs used our measurement.

\begin{figure}[htbp]
 \begin{center}
  \includegraphics[width=30mm,bb=0 0 256 227]{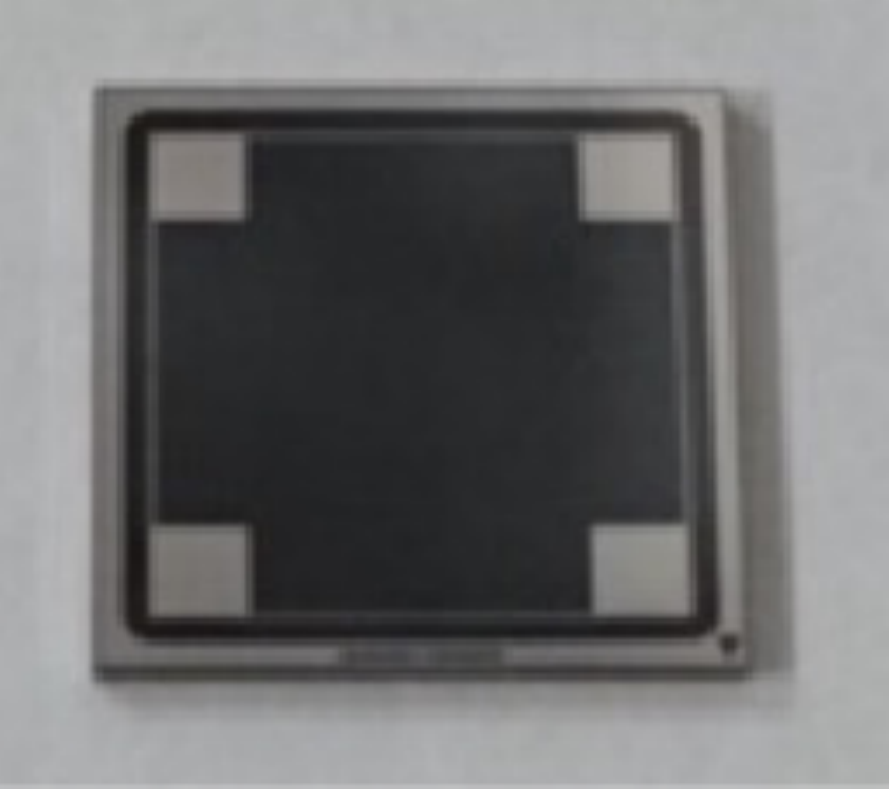}
 \end{center}
 \caption{The PSD sensor.  }
 \label{fig:PSDsensor}
\end{figure}

Figure \ref{fig:PSDsensor} shows a PSD sensor made by Hamamatsu Photonics.
This sensor has electrodes at the four corners.

\begin{figure}[h]
%	\centering
	\begin{subfigure}{0.4\columnwidth}
		\centering
		\includegraphics[width=30mm,height=40mm,keepaspectratio,clip,bb=0 0 3264 2448]{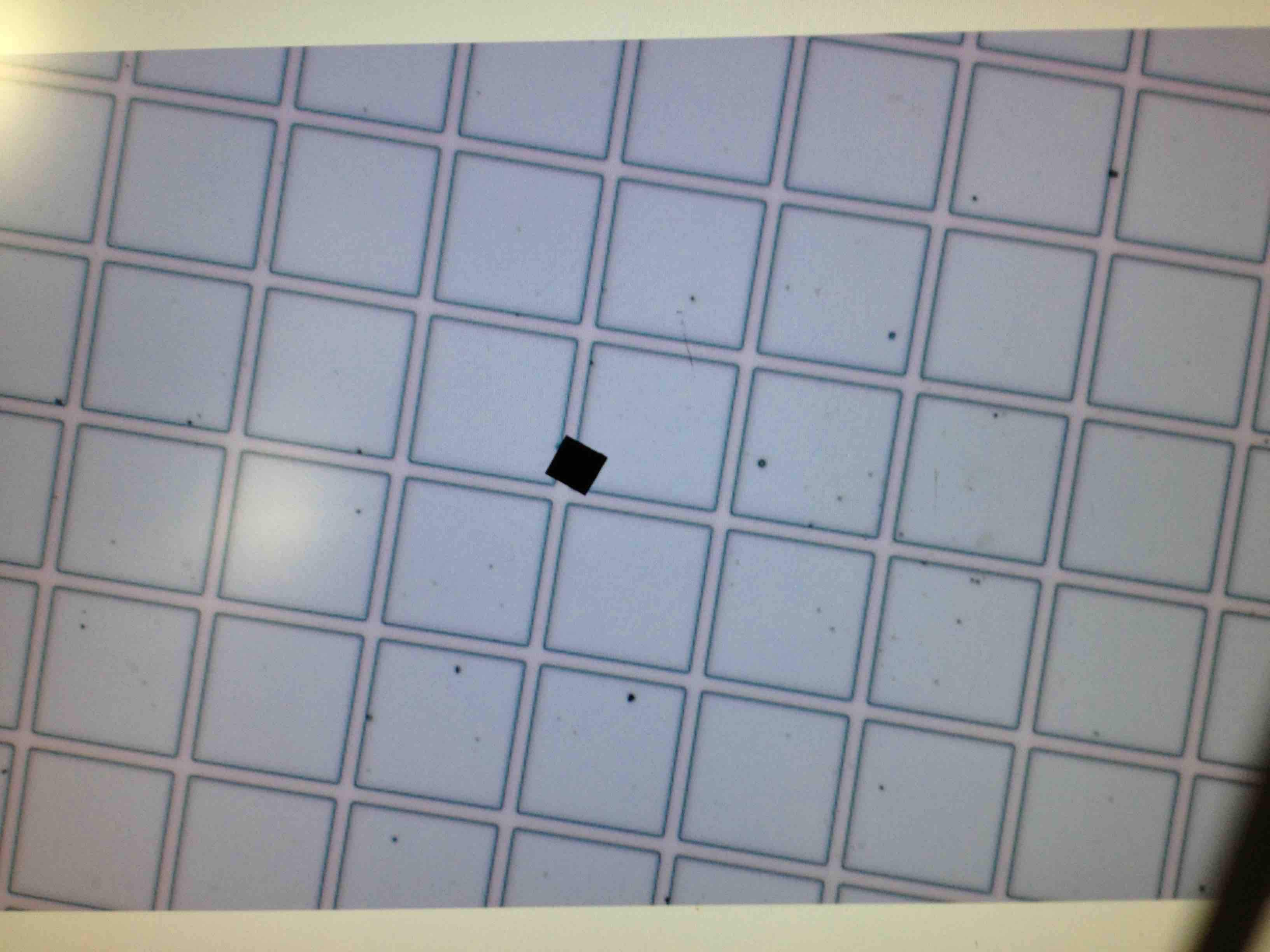}
		\caption{meshed p$^+$ layer}
		\label{fig:mesh}
	\end{subfigure}
	\centering
	\begin{subfigure}{0.4\columnwidth}
		\centering
		\includegraphics[width=30mm,height=40mm,keepaspectratio,clip,bb=0 0 3264 2448]{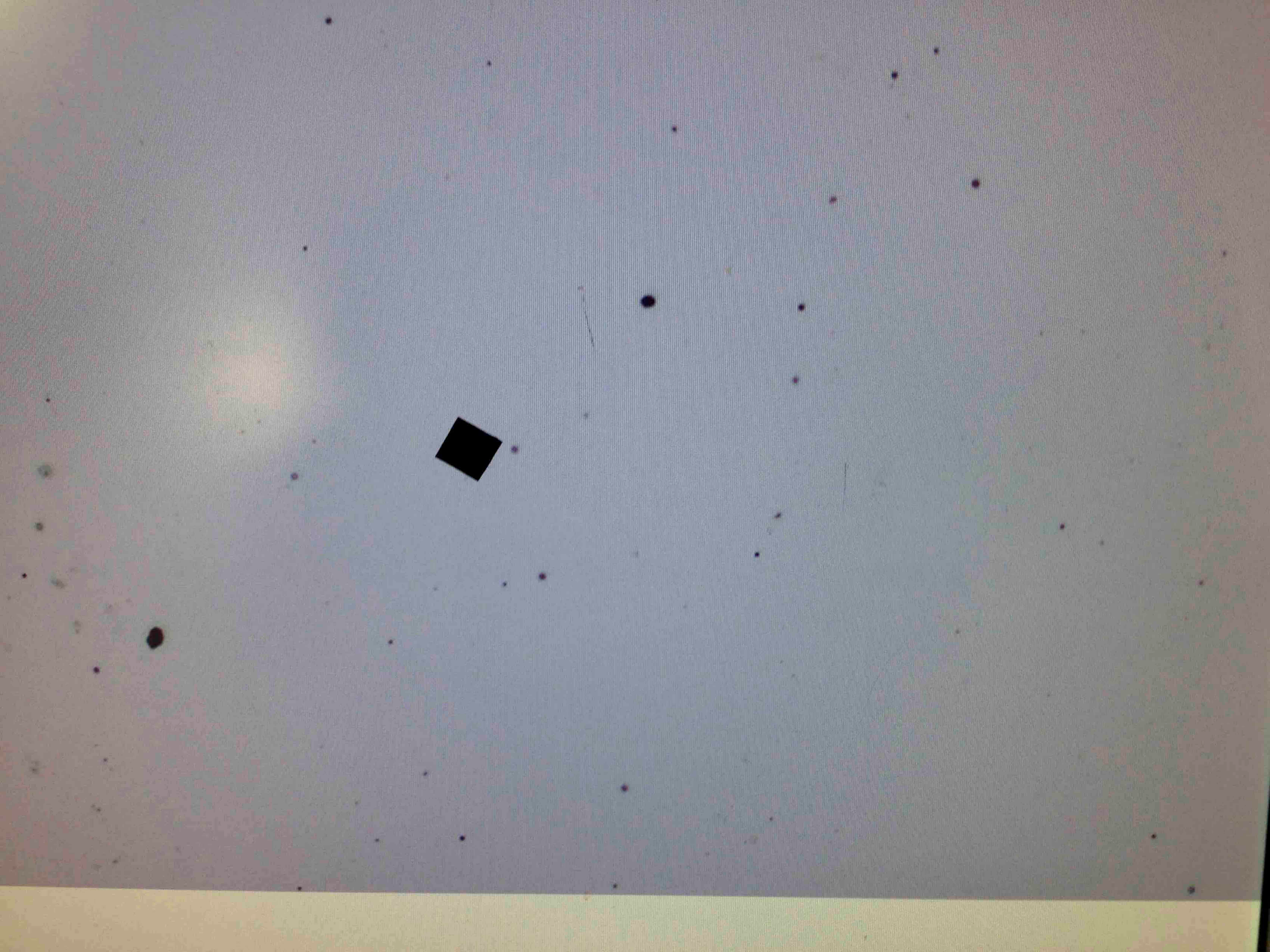}
		\caption{non-meshed p$^+$ layer}
		\label{fig:nonmesh}
	\end{subfigure}
	\caption{magnified picture of PSD }
	\label{fig:magpsd}
\end{figure}

We have two types of PSD sensors. The difference between the two types can be seen by a microscope.
Figure \ref{fig:mesh} and \ref{fig:nonmesh} are the magnified views of the black areas of Fig.3 for the two types of sensors. %Fig\ref{fig:nonmesh} has mesh, but Fig.\ref{fig:mesh} has mesh on the P+ surface. 
This mesh increases the resistivity of the p$^+$ layer. With the larger resistivity, it is expected to reduce the noise and the position distortion. 
Flat surface of the p$^+$ layer is seen on Figure \ref{fig:nonmesh}, in contrast to Figure \ref{fig:mesh}, which shows meshed p$^+$ surface.

\begin{figure}[h]
	\centering
	\begin{subfigure}[b]{0.4\columnwidth}
		\centering
		\includegraphics[width=\columnwidth,keepaspectratio,angle=270,bb=5 117 590 714]{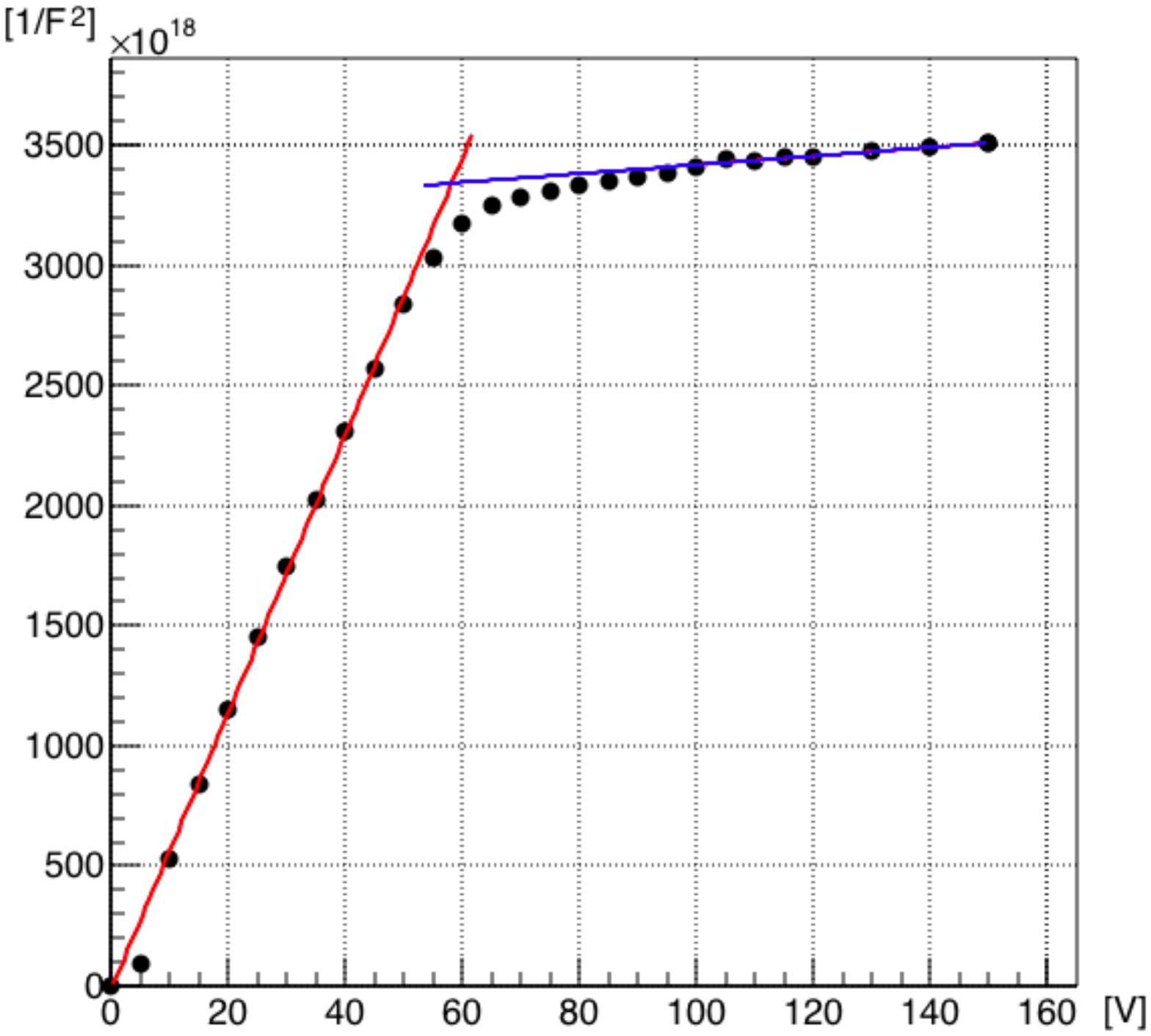}
		\caption{meshed PSD}
		\label{fig:abcdddd}
	\end{subfigure}
	\centering
	\begin{subfigure}[b]{0.4\columnwidth}
		\centering
		\includegraphics[width=\columnwidth,keepaspectratio,angle=270,bb=7 102 588 728]{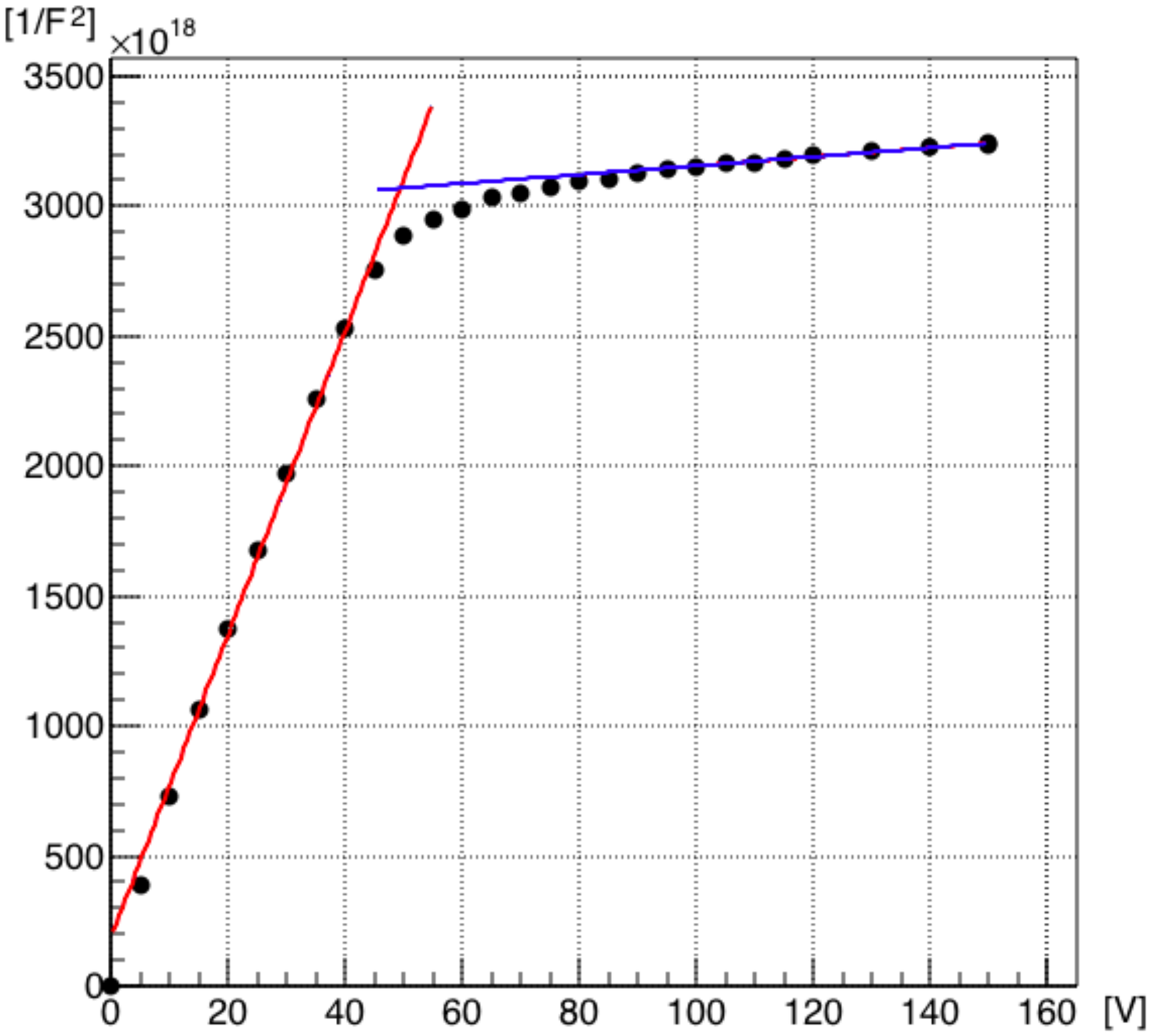}
		\caption{non-meshed PSD}
		\label{fig:abcdd}
	\end{subfigure}
	\caption{C-V characteristics }
	\label{fig:cv} 
\end{figure}

Figure \ref{fig:abcdddd} and \ref{fig:abcdd} are the result of capacitance measurement characteristics of PSD with mesh and PSD without mesh, respectively. 
 They are fully depleted at about 60 V and 50 V, respectively.
 
\begin{figure}[htbp]
\centering
 	\begin{minipage}[b]{0.4\columnwidth}
		\centering
  \includegraphics[width=50mm,keepaspectratio,bb=0 0 1007 755]{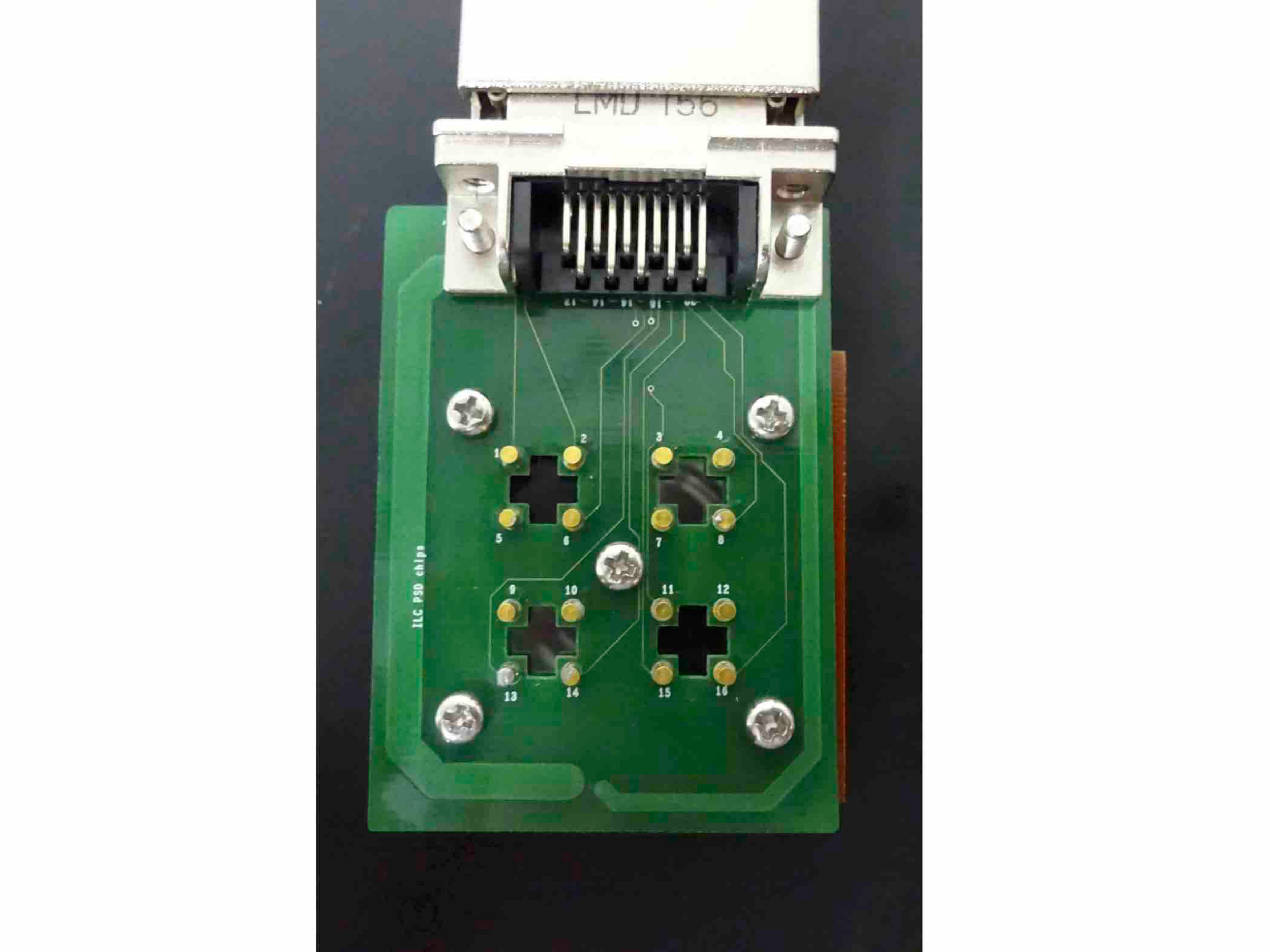}
  		\label{fig:abcd}
 \caption{Sensor Box}
  \label{fig:sensorbox}
 	\end{minipage}
\begin{minipage}[b]{0.4\columnwidth}
 \begin{center}
  \includegraphics[width=50mm,keepaspectratio,bb=0 0 457 343]{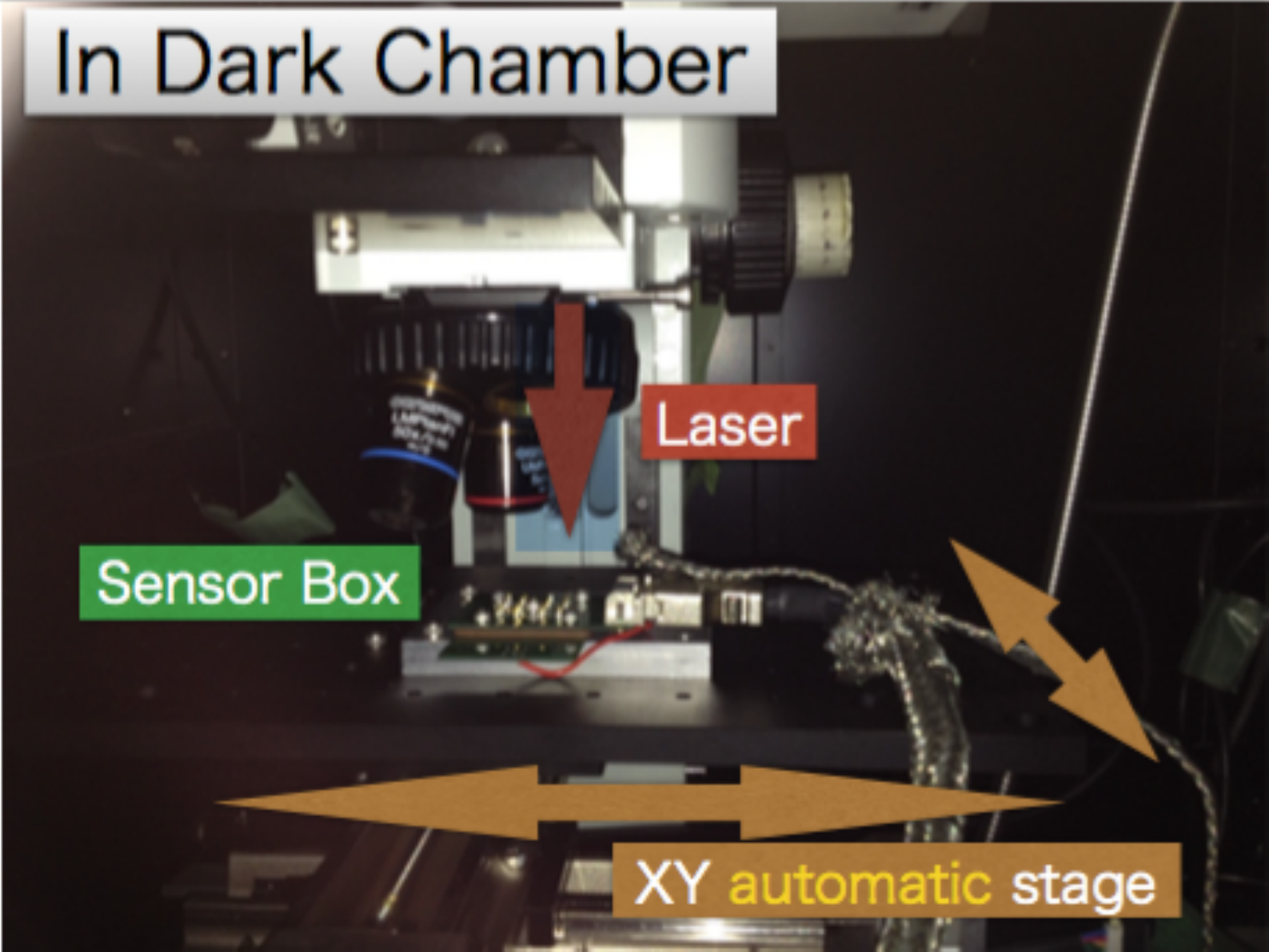}
 \end{center}
 \caption{Sensor Box fixed on XY stage}
 \label{fig:XYstage}
 	\end{minipage}
\end{figure}

Figure \ref{fig:sensorbox} is a printed circuit board (PCB) and a holder for four sensors. This is called ``sensor box" below.  In this picture two of the four sensor places are filled with a meshed and a non-meshed sensor. The box was fixed in a two axis automatic stage in a dark chamber as shown in Figure \ref{fig:XYstage} and reverse bias of 100 V was applied to the sensors.

\section{Evaluation of Detecting Position using Laser}

Laser photons were injected to the PSD sensors to test the position reconstruction. 
The specifications of the laser are shown in Table \ref{tab:laser}.

\begin{table}[h]
\begin{center}
  \begin{tabular}{|c|c|} \hline
    Type & GryLaS IQ-1064-2 \\ \hline
    Wavelength　& 1064 nm \\ \hline
    Laser intensity  &  20 $\mu$J, reduced by NID filiters to 50 MIP equivalent.   \\   \hline
    Pulse duration &1.2 ns\\  \hline
    Repetition rate & 1000 Hz \\ \hline
    Laser spot size & less than 20 $\mu$m  \\ \hline
  \end{tabular}
\end{center}
\caption{The specifications of the laser and optics}
\label{tab:laser}
\end{table}

The PCB on the sensors has cut on the main part of PSD sensors to pass the laser photons through it.
Since the photon energy is slightly higher than the band gap energy of silicon, one optical photon creates one electron-hole pair in the silicon, and imitate signal at the well defined position. The movable stage below the sensor box was used to control the injection position.

%As a Fig.\ref{fig:XYstage}, laser light equivalent to 50 MIP is incident on various positions of PSD sensor. The laser used was 1000 nm in wavelength. This laser behaves similarly to incident particles in real machines. Laser light was applied at 1000 Hz for 8 seconds for each point. Then, the position is reconstructed from the charge measured from each electrode.

\begin{figure}[h]
%\begin{minipage}{0.4\columnwidth}
 \centering
  \includegraphics[width=80mm]{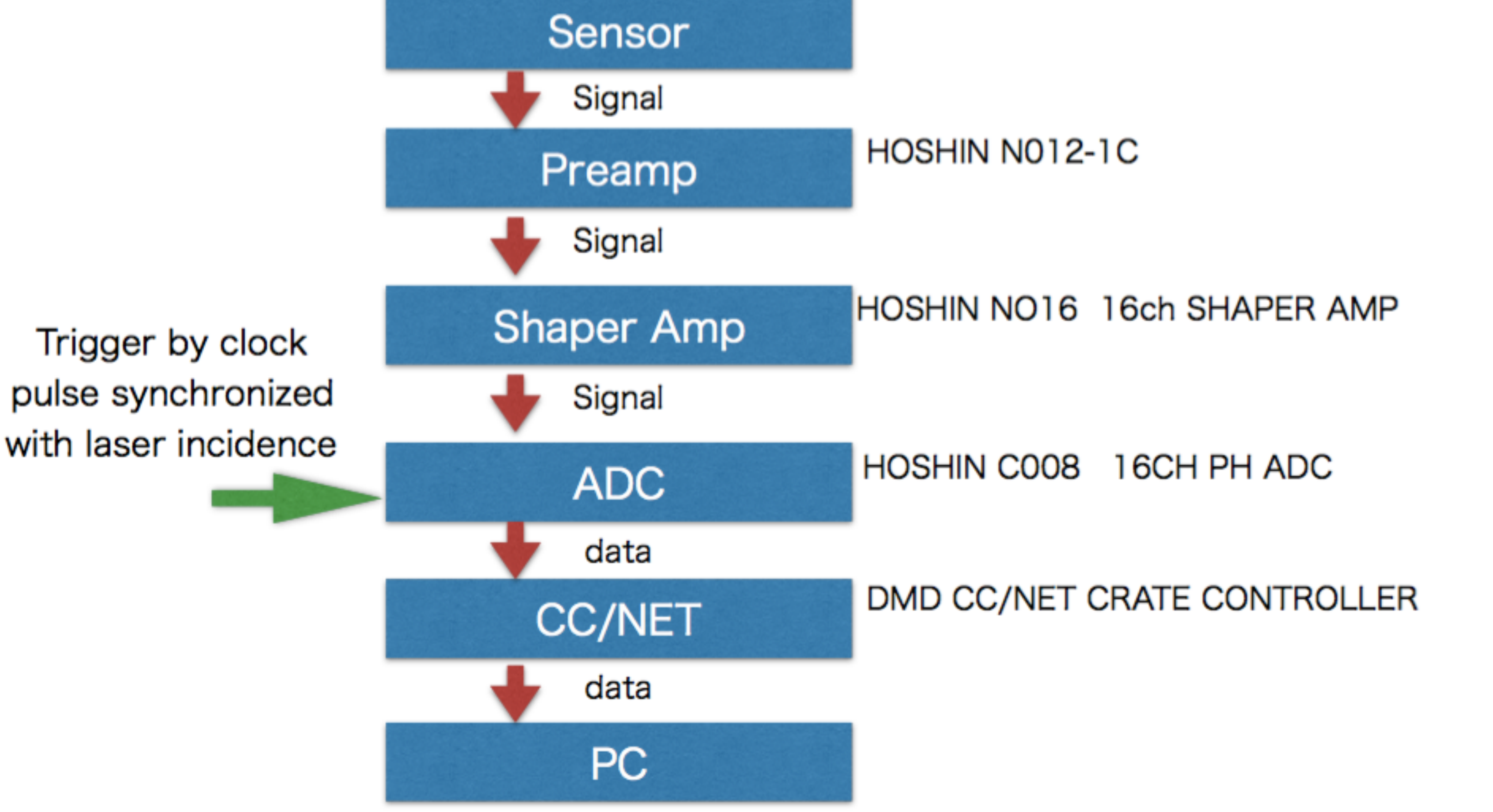}
 \caption{A schematic diagram of the DAQ}
 \label{fig:ffff}
\end{figure}

\begin{figure}
 \centering
  \includegraphics[width=45mm,bb=-100 0 450 250]{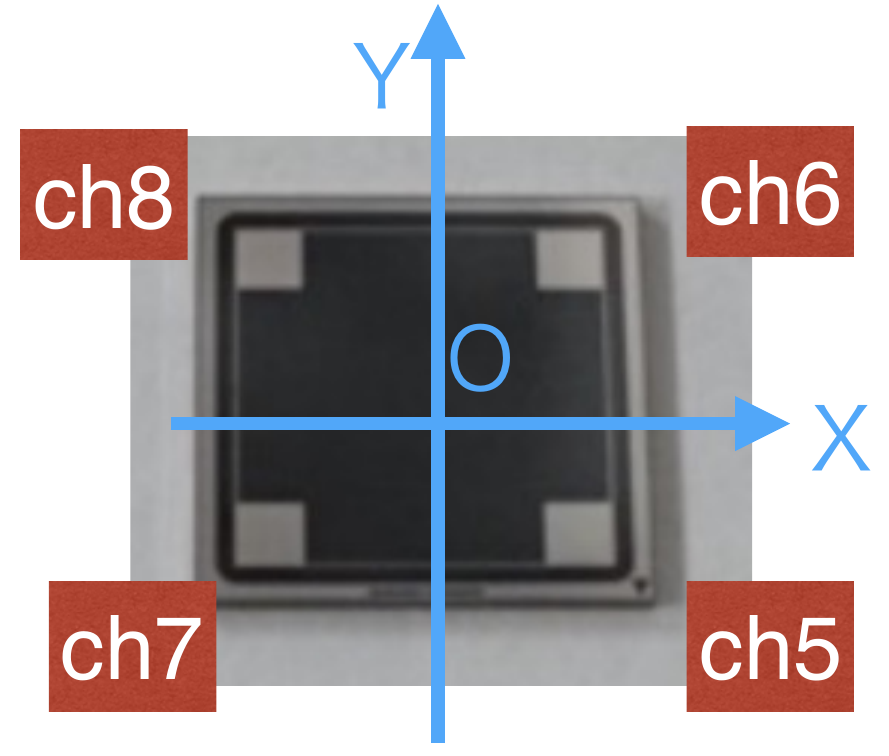}
 \caption{Assignment of channels on the PSD electrode.}
 \label{fig:daxis}
% \end{minipage}
\end{figure}

Figure \ref{fig:ffff} shows a schematic diagram of data  acquisition system (DAQ). 
Signal from each electrode was amplified by a preamplifier and a shaper, and delivered to a peak-hold ADC module on a CAMAC system.
The gate signal of the ADC was applied from the laser injection trigger.
We accumulate 8000 pulses at one point, and average the ADC counts over the pulses. 
To calculate the position from signals, we use the following formulae \cite{PSDmath},

\begin{gather}
X_{\mathrm{rec}}=\frac{(\mathrm{ch5+ch6})-(\mathrm{ch7+ch8})}{\mathrm{ch5+ch6+ch7+ch8}}              \\
Y_{\mathrm{rec}}=\frac{(\mathrm{ch6+ch8})-(\mathrm{ch5+ch7})}{\mathrm{ch5+ch6+ch7+ch8}}  \notag 
\label{equ:rec}
\end{gather}

where $X_{\mathrm{rec}}$ and $Y_{\mathrm{rec}}$ are reconstructed position along $X$ and $Y$ axis, and ch5-8 stand for ADC counts after the pedestal subtraction, as shown in Fig. \ref{fig:daxis}.

%\end{minipage}
%\begin{minipage}{0.4\columnwidth}

%As I mentioned earlier, we use the ratio of the charges collected at the electrode depending on the arrival position.
%For example, I named each electrodes channel 5, 6, 7 and 8, respectively. Then I defined the axis like this plot. The X and Y positions can be calculated with these formulae. Each ch of these formulae means ADC output from each ch after the pedestal subtraction.

\subsection{Measurement result of meshed PSD}

\begin{figure}[htbp]
\centering
\begin{subfigure}[b]{0.4\columnwidth}
\centering
  \includegraphics[width=50mm,height=40mm,keepaspectratio,bb=0 0 530 475]{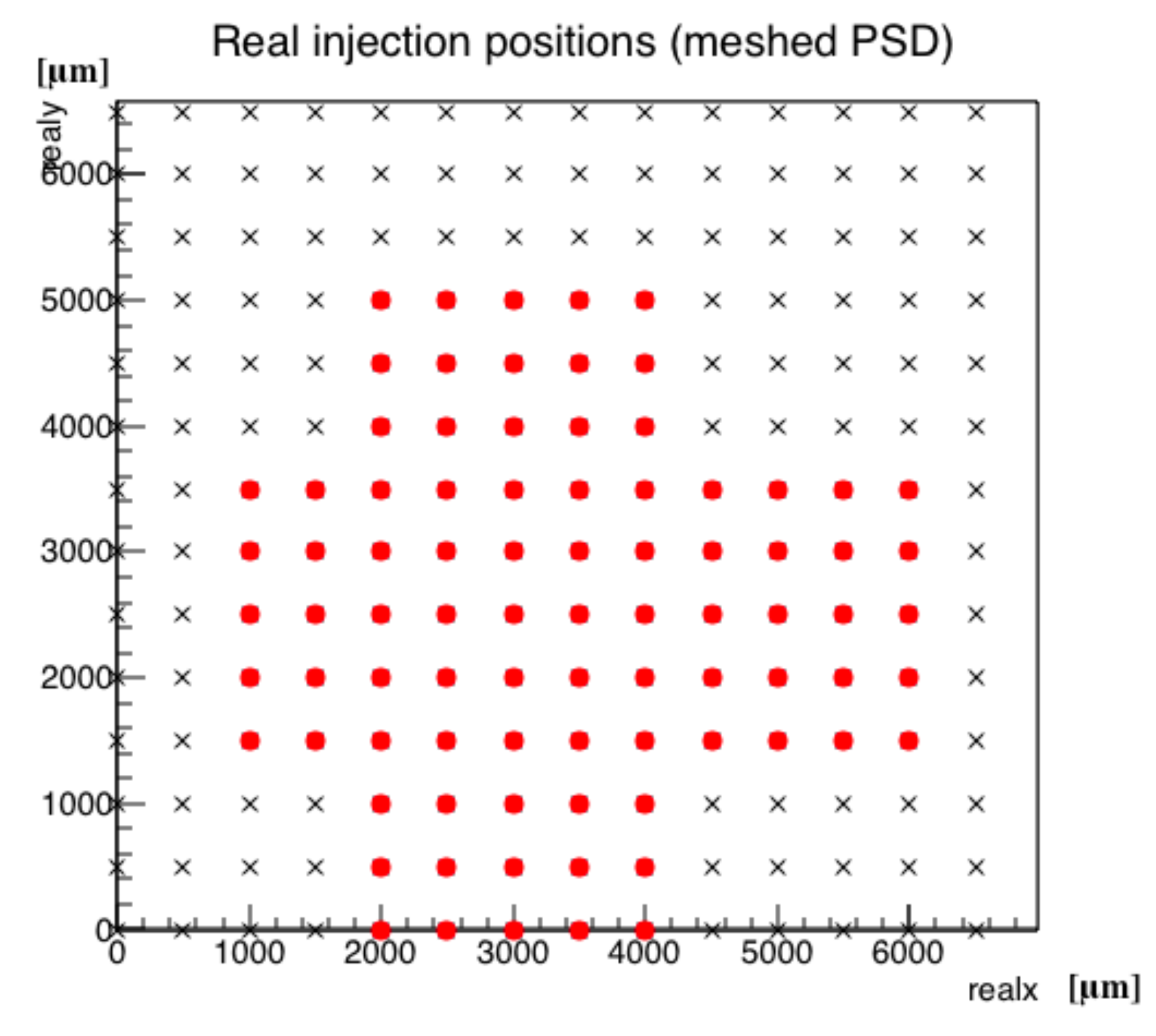}
 \caption{Real injection positions}
 \label{fig:realmesh}
\end{subfigure}
\begin{subfigure}[b]{0.4\columnwidth} \begin{center}
\centering
  \includegraphics[width=50mm,height=40mm,keepaspectratio,bb=0 0 698 475]{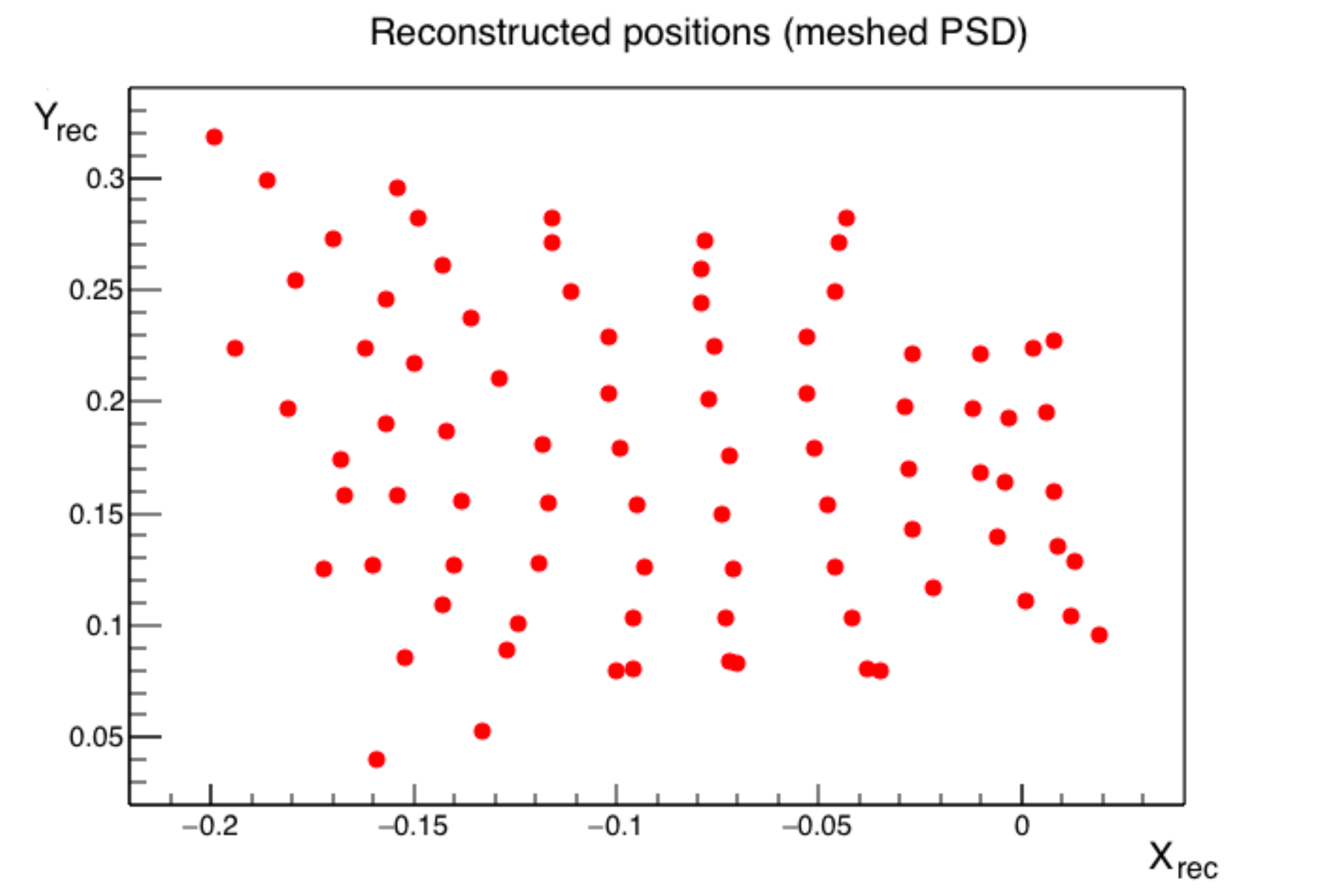}
 \end{center}
 \caption{Reconstructed positions}
 \label{fig:recmesh}
 \end{subfigure}
 	 \caption{Measurement result of meshed PSD}
 \label{fig:meshed}
\end{figure}

Figure \ref{fig:realmesh} shows real injection positions, obtained from the positions of the two-axis stage. The red points are  the point with sufficiently strong signals of 1000 or more ADC counts on all channels. A black cross mark indicates that the ADC output value of one or more channels is smaller than 1000 so that the signal is too weak to be reconstructed.
This plot should reflect the shape of the open cut of the sensor box if the sensor is fully active.

Figure \ref{fig:recmesh} shows reconstructed positions by Eq.1. In this figure, only points with sufficient signal strength are plotted.
Distortion from the original grid, which is expected behavior of PSDs, is seen.

\begin{figure}[h]
	\centering
	\begin{subfigure}{0.4\columnwidth}
		\centering
		\includegraphics[width=\columnwidth,bb=0 0 698 475]{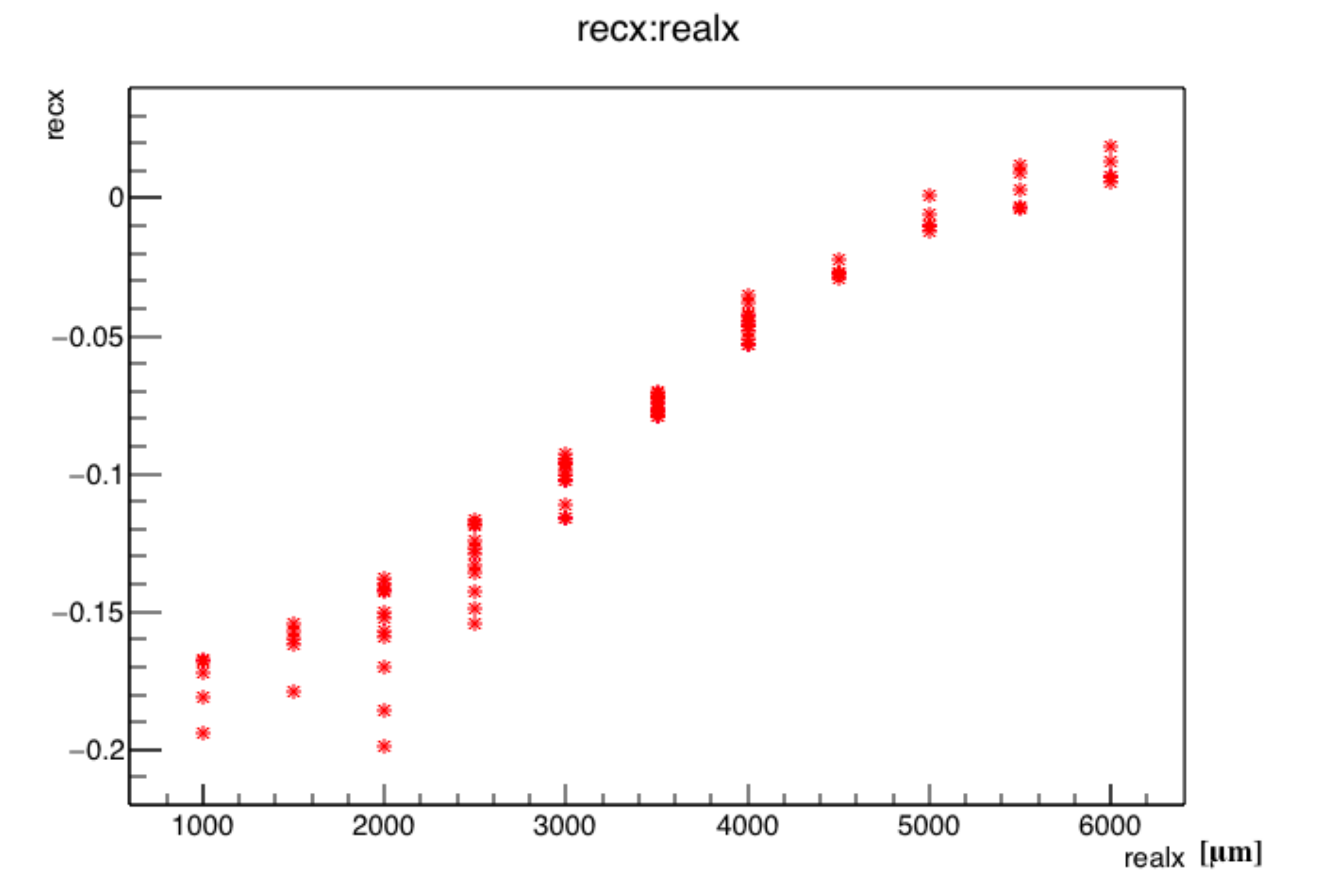}
		\caption{X coordinates}
		\label{fig:meshcolx}
	\end{subfigure}
	\begin{subfigure}{0.4\columnwidth}
		\centering
		\includegraphics[width=\columnwidth,bb=0 0 698 475]{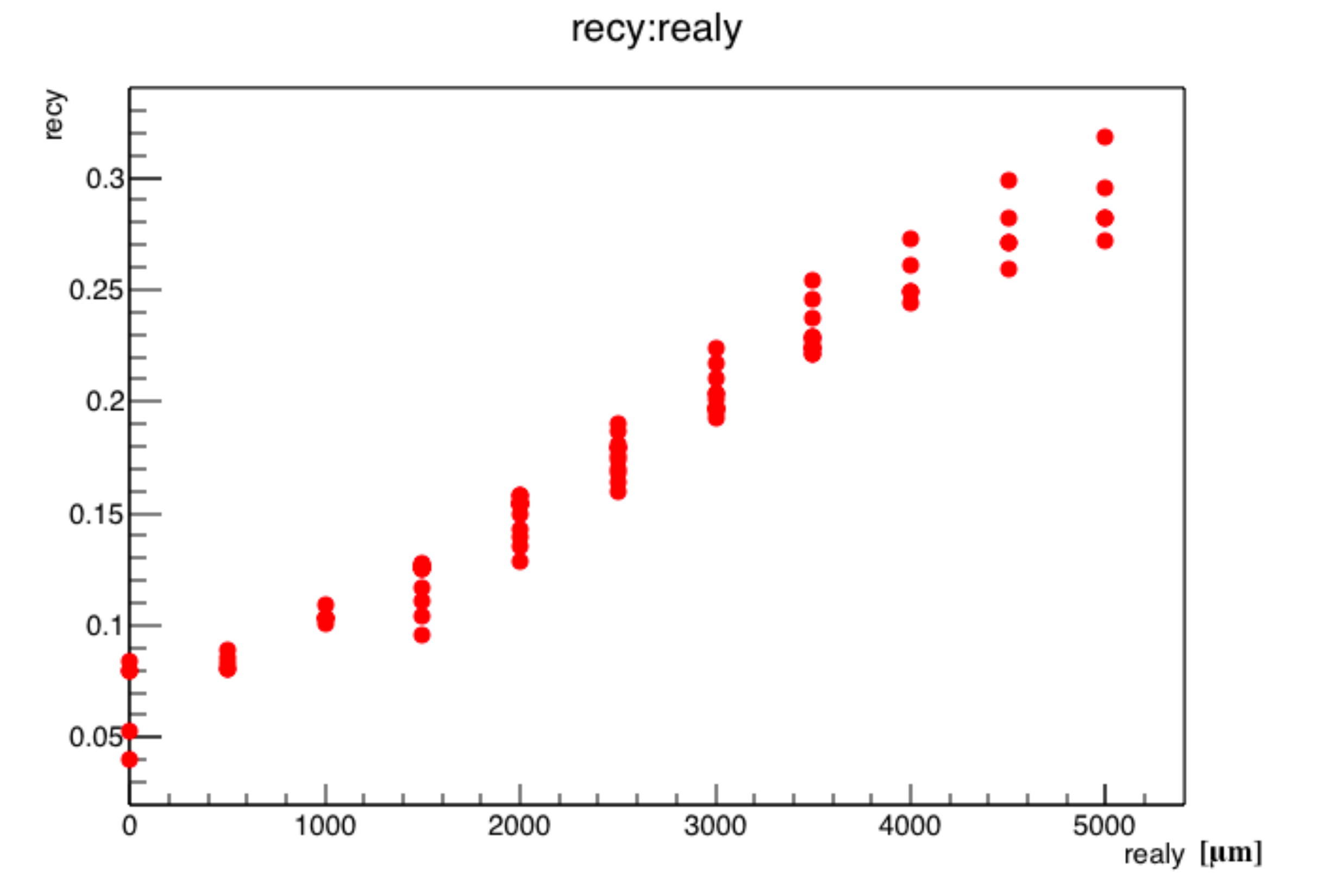}
		\caption{Y coordinates}
		\label{fig:meshcoly}
	\end{subfigure}
	 \caption{The correlation result of meshed PSD}
 \label{fig:col}
\end{figure}

Figure \ref{fig:meshcolx} and \ref{fig:meshcoly} are the correlation between the actual incident position and the reconstructed position of the X and Y coordinates, respectively. A good correlation between the two positions with some non-uniformity is obtained.

\subsection{Measurement result of non-meshed PSD}

\begin{figure}[htbp]
\centering
\begin{subfigure}{0.4\columnwidth}
 \begin{center}
  \includegraphics[width=59mm,keepaspectratio,bb=0 0 571 475]{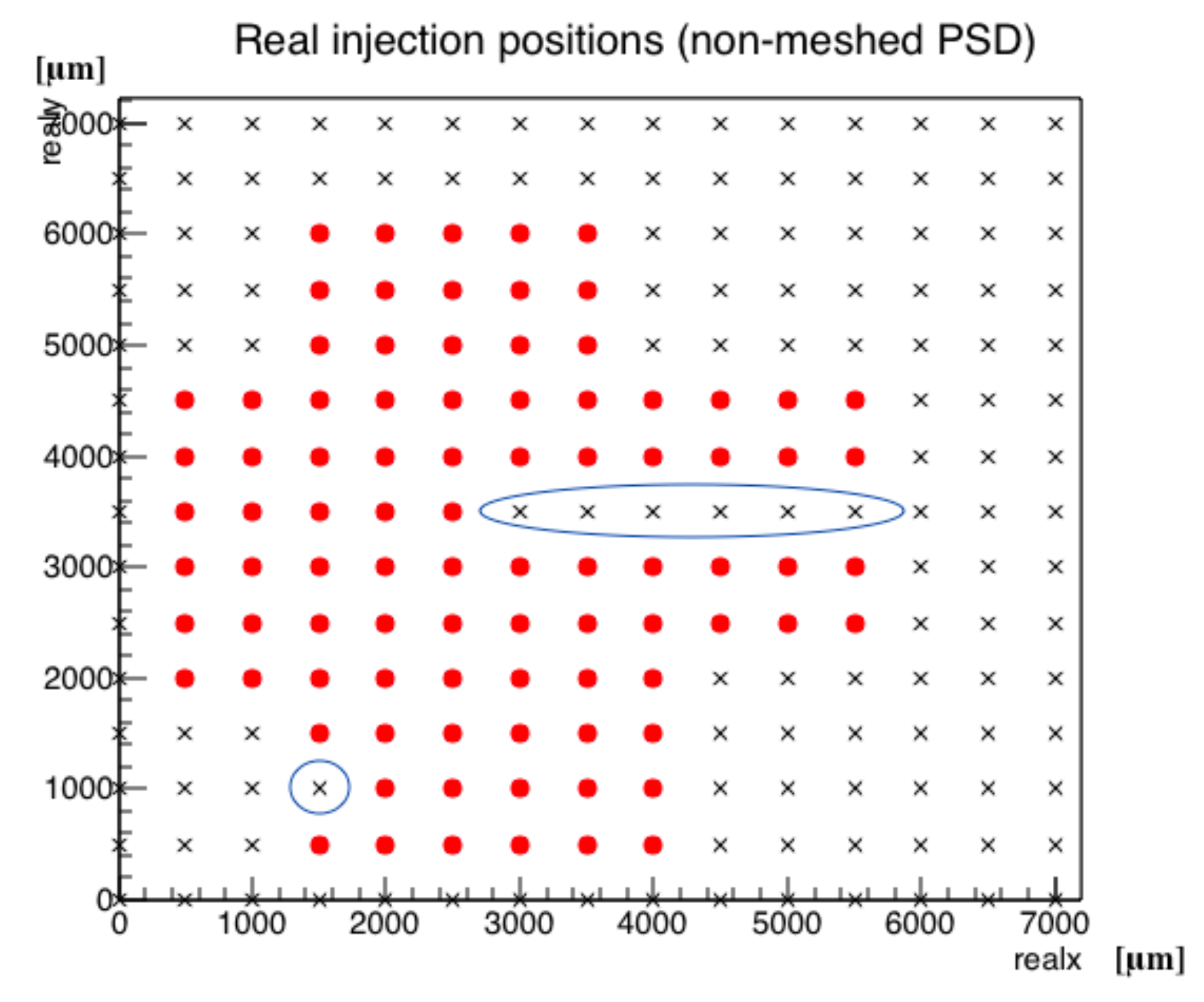}
 \end{center}
 \caption{Real injection positions}
 \label{fig:realnonmesh}
\end{subfigure}
\begin{subfigure}{0.4\columnwidth}
 \begin{center}
  \includegraphics[width=71mm,keepaspectratio,bb=0 0 864 584]{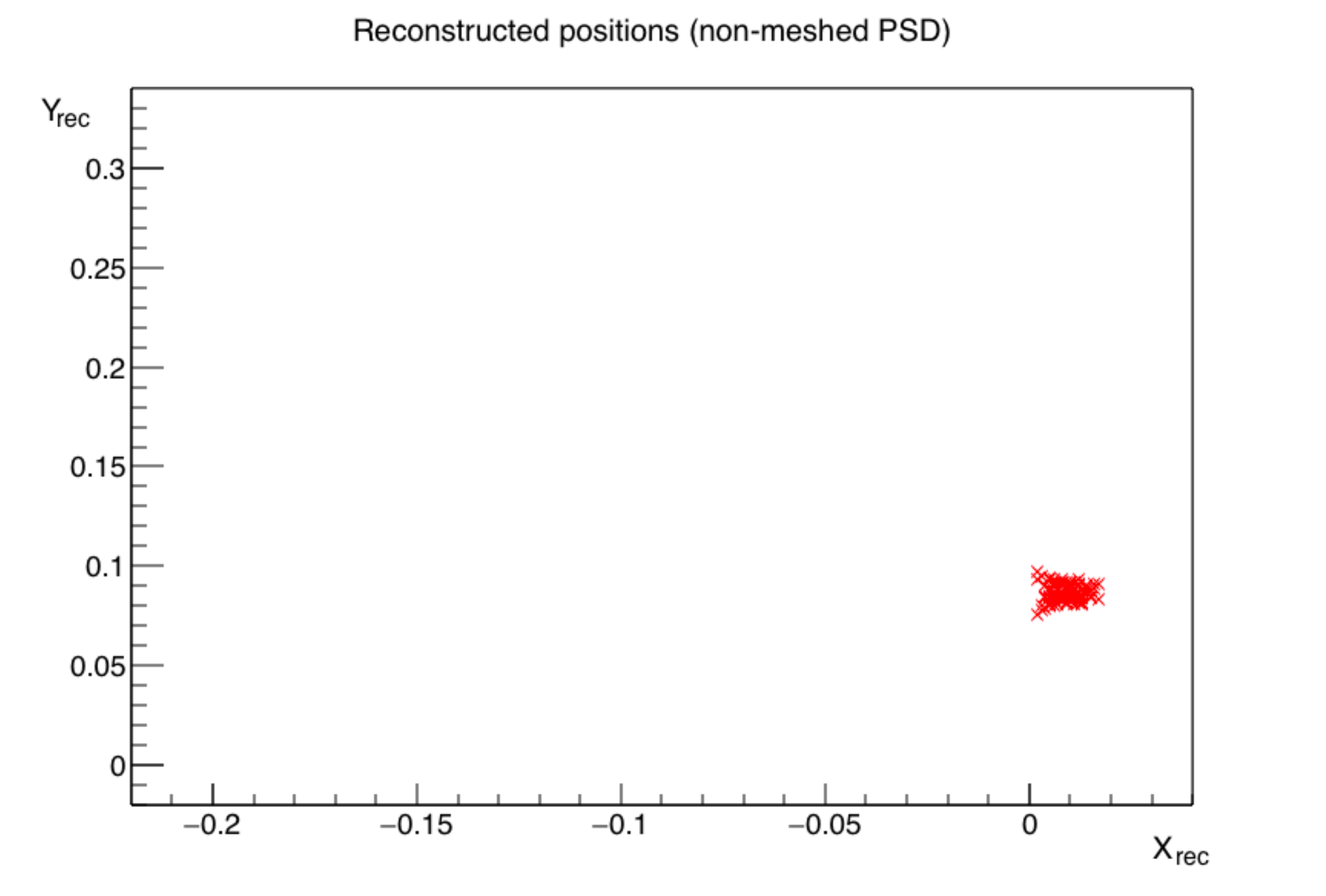}
 \end{center}
 \caption{Reconstructed positions}
 \label{fig:recnonmesh}
 \end{subfigure}
 \caption{Measurement result of non-meshed PSD}
 \label{fig:non-m}
\end{figure}

%Figure \ref{fig:realnonmesh} shows real injection positions, obtained from the positions of the two-axis stage. 
The same measurement was performed on the non-meshed PSD, shown in figure \ref{fig:realnonmesh}.
%Red points and black cross points has the same mean as the meshed PSD, and
 Shapes like open cut of sensor box are seen. 
However, strong signals cannot be obtained at two places surrounded by an ellipse, and the reason is under investigation.

Figure \ref{fig:recmesh} shows reconstructed positions. Compared with the meshed PSD, the reconstructed positions are concentrated in the almost same value.

This shows that the non-meshed PSDs are not functional to obtain the incident positions.
The high resistivity of p$^+$ layer should be essential for this type of PSDs.
%Fig\ref{fig:nonmeshcolx} and fig.\ref{fig:nonmeshcoly} are the correlation between the actual incidence position and the reconstructed position of the X and Y coordinates respectively. Compared with the meshed PSD, it can be seen that there is almost no correlation.

%\begin{figure}[h]
%	\centering
%	\begin{minipage}{0.4\columnwidth}
%		\centering
%		\includegraphics[width=70mm,height=40mm,keepaspectratio,clip]{002colx-7743.png}
%		\caption{The correlation of X coordinates}
%		\label{fig:nonmeshcolx}
%	\end{minipage}
%	\centering
%	\begin{minipage}{0.4\columnwidth}
%		\centering
%		\includegraphics[width=70mm,height=40mm,keepaspectratio,clip]{002coly-7718.png}
%		\caption{The correlation of Y coordinates}
%		\label{fig:nonmeshcoly}
%	\end{minipage}
%\end{figure}

%From these things, it is thought that  high-resistivity of P+ layer is essential.

\section{A noise measurement using $\beta$ sourse}

\begin{figure}[h]
 \begin{center}
  \includegraphics[width=50mm,bb=0 0 388 527]{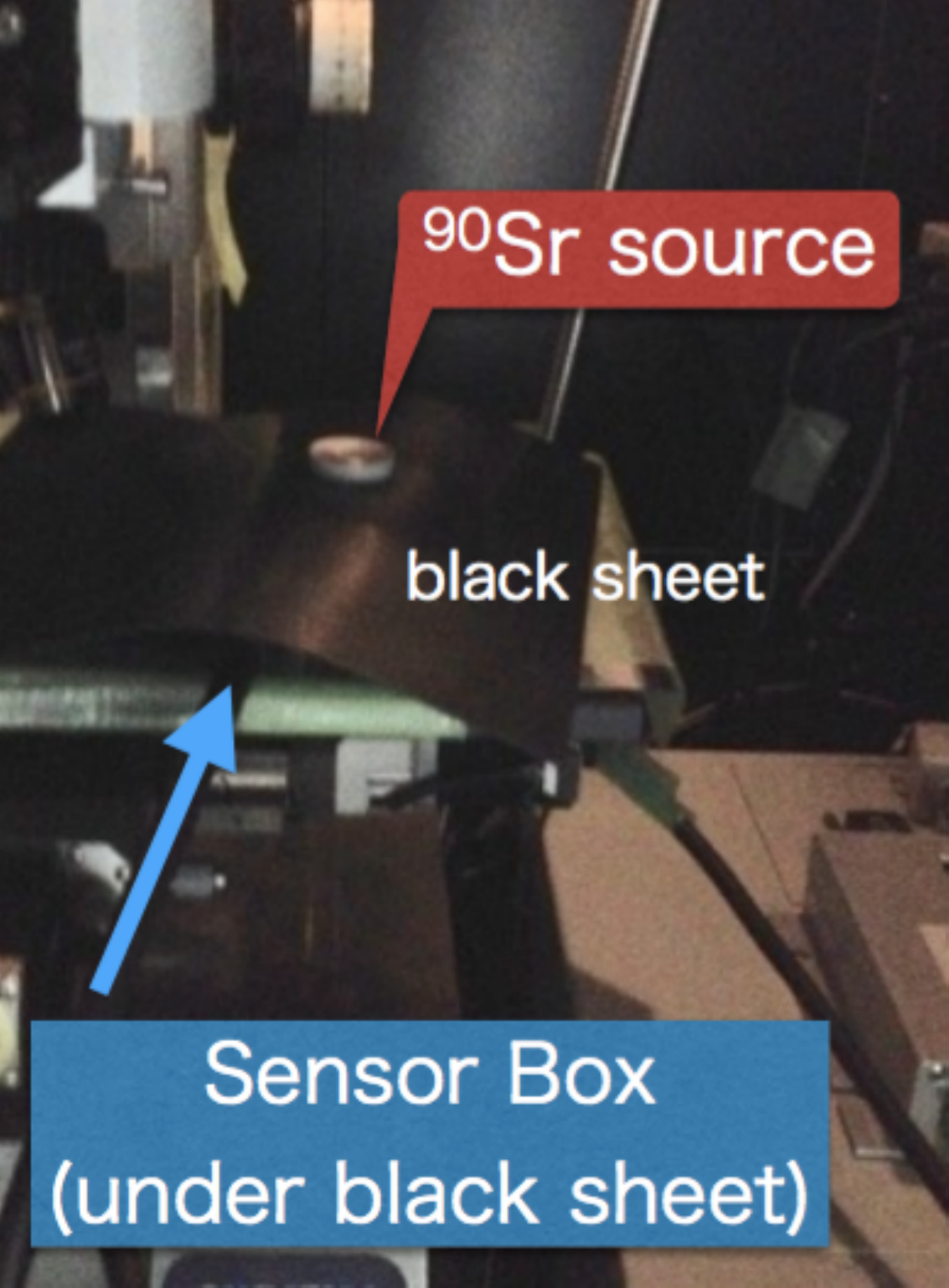}
 \end{center}
 \caption{Radiation source put in the Sensor box}
 \label{fig:rstage}
\end{figure}

As shown in Figure \ref{fig:rstage}, a rubber sheet of 1 mm thickness was put on the sensor box, and $\mathrm{^{90}Sr}$ was put on the rubber sheet. That was done with a conventional pixel type silicon sensor and PSD with mesh, respectively, and we tried to capture the signal with the radiation source.

\begin{figure}[h]
 \begin{center}
  \includegraphics[width=80mm,bb=0 0 947 519]{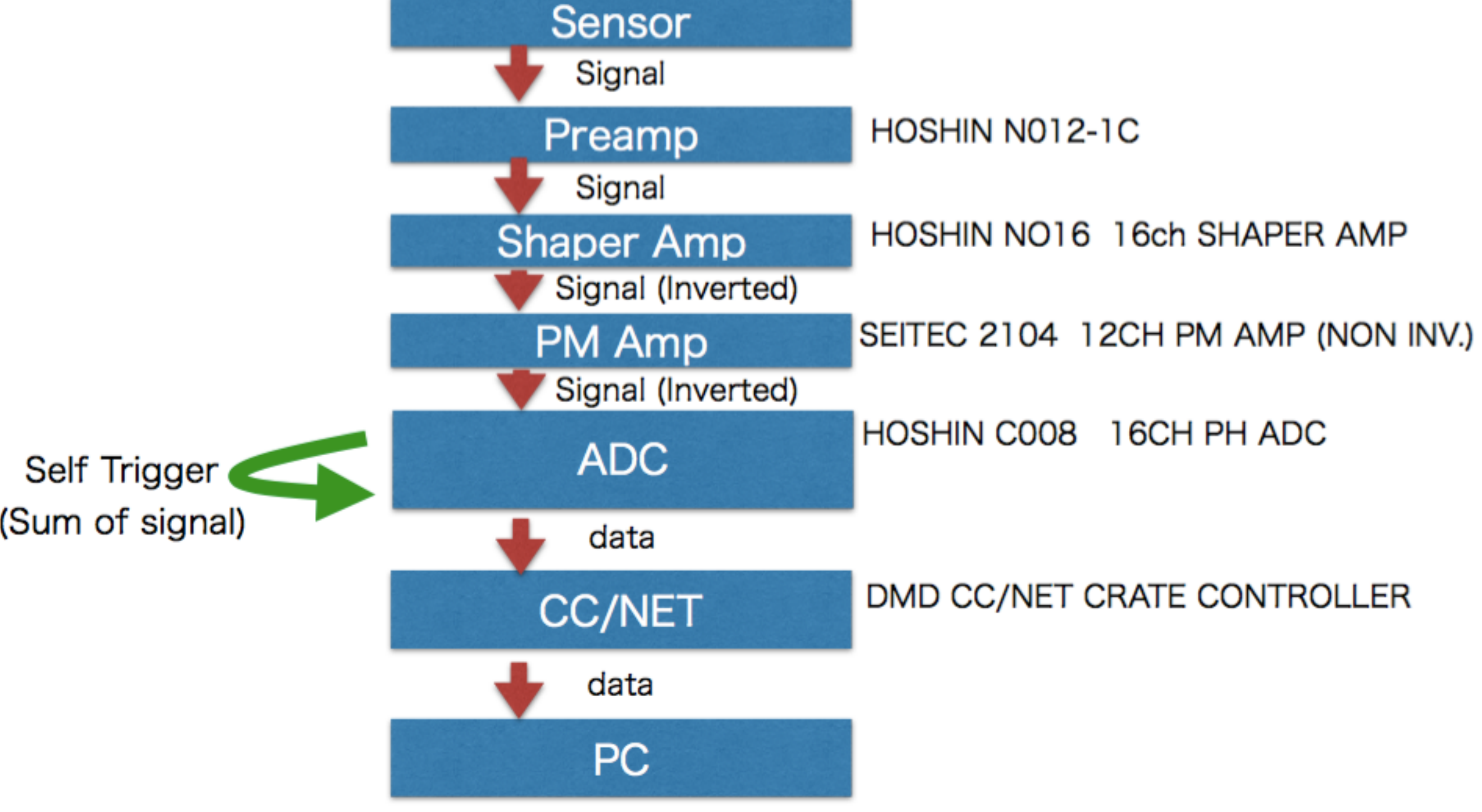}
 \end{center}
 \caption{A schematic diagram of DAQ to capture the signal by the radiation source}
 \label{fig:gggg}
\end{figure}

Figure \ref{fig:gggg} shows a schematic diagram of DAQ to capture the signal with the radiation source.  ADC was self-triggered. Signals are inverted in the shaper amplifier to make the trigger.
For the PSD sensor, measurement was carried out for 2 hours in a state with and without a radiation source. During this time, the sum of the signals from four electrodes after passing shaper amplifier from the four electrodes exceeded 400 mV was treated as an event.
In the pixel type silicon sensor, 4 pixels out of 9 pixels were used in a state with a radiation source. When the signal of 1 pixel exceeded 100 mV, it was treated as an event. At this time, the output values of the other three pixels are recorded as a pedestal.

\begin{figure}[htbp]
 \centering
 %\begin{subfigure}{0.4\columnwidth}
  \includegraphics[width=80mm,bb=0 0 698 475]{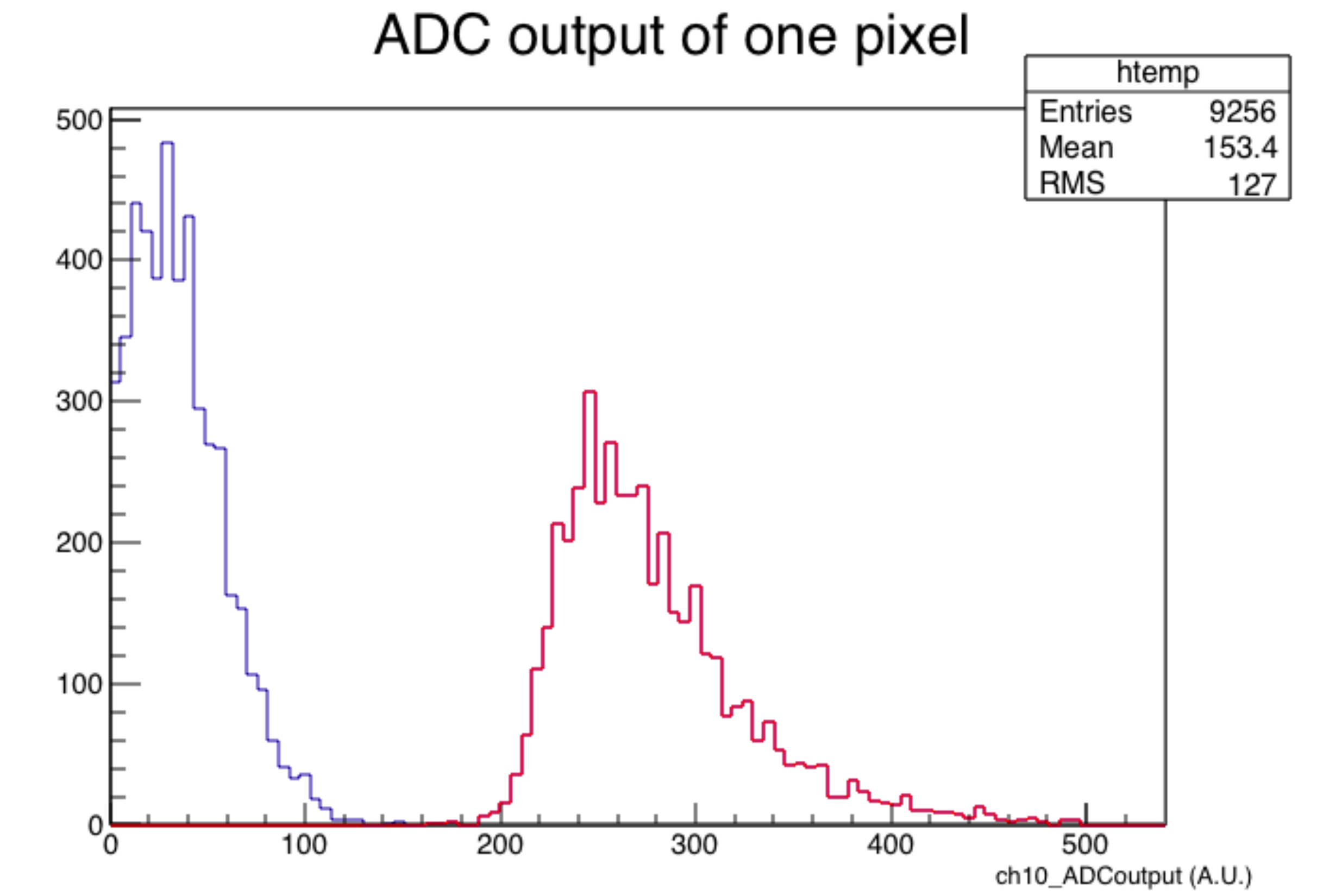}
 \caption{The distribution of the ADC output in one pixel of a conventional pixel type silicon sensor. (Red: Signal from radiation source  Blue: Pedestal)}
 \label{fig:baby44}
\end{figure}

Figure \ref{fig:baby44} is the distribution of the ADC output in one pixel of a conventional pixel type silicon sensor. It seems that the signal from the beta source on the right side of the plot and the pedestal on the left seem to be separated well, but it seems to be seen separately by the trigger.

\begin{figure}[htbp]
%\begin{minipage}{0.4\columnwidth}
\centering
  \includegraphics[width=80mm,bb=0 0 698 475]{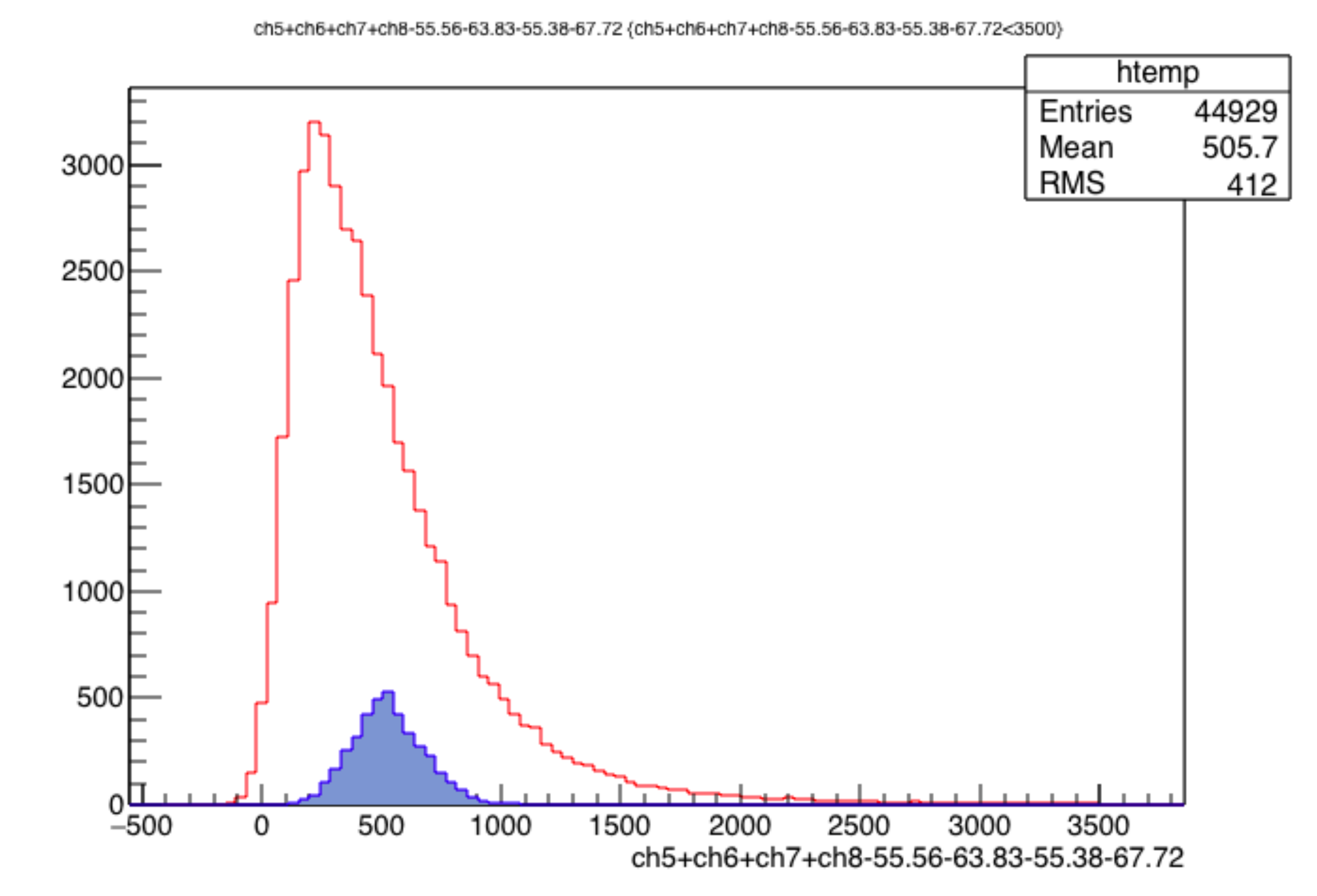}
 \caption{The distribution of the sum of the ADC outputs from the four electrodes of PSD (Red: with radiation source  Blue: without radiation source )}
 \label{fig:aaaa}
 %\end{minipage}
\end{figure}

Figure \ref{fig:aaaa} is the distribution of the sum of the ADC outputs from the four electrodes of PSD with and without radiation source. This shows that the signal comes more frequently with radiation source than without, however the difference on the distribution is not clear, mainly due to a noise coherent to all channels.
%The red histogram shows the distribution in the case with the radiation source and the blue distribution shows the case without the radiation source. It is measured acqu the same length of, since the number of events in the case with the radiation source is obviously larger than in the case without the radiation source, the signal from the radiation source is considered to be visible in the PSD. However, it turns out that the coherent noise of each channel is too large to separate from the signal. 
We plan to reduce this system noise in the future and try again to measure the radiation source.

%\end{minipage}

\section{Summary}
PSD is a silicon device which can derive the incident position of a particle by resistive devision of the charge to electrodes at the p$^+$ surface.  It is expected PSDs at the innermost layers of ECAL to improve the position resolution of photons. Our first sample shows reasonable reconstruction of incident position of laser photons with some distortion. Meshed p$^+$ surface gives better result. Studies on the noise is ongoing.

The silicon pads measured in this study have no gain. Silicon sensors with avalanche gain are recently developed  and is expected to obtain position resolution less than 100 $\mathrm{\mu m} $.  We plan to reduce electronic noise for PSD measurement, create PSD sensors with avalanche gain, and confirm effects on physics performance with Monte-Carlo simulation study.

\section*{Acknowledgements}
We appreciate that J-PARC muon g-2/EDM collaboration supported in production of the PSD sensors, and Hamamatsu Photonics  suggested meshed PSD sensor.
%Thank you for making PSD sensor using the production margin of J-PARC muon g-2 / EDM experiment.
%Thank you for suggesting meshed PSD sensor for Hamamatsu Photonics.

\end{document}